\documentclass[english,reprint,showpacs]{revtex4-1}  
\usepackage{dcolumn}   
\usepackage{xcolor}
\usepackage{mathptmx}
\usepackage[T1]{fontenc}
\usepackage[latin9]{inputenc}
\usepackage{color}
\usepackage{amsmath}
\usepackage{amssymb}
\usepackage{hyperref}
\usepackage{graphics}
\usepackage{graphicx}
\usepackage{epsfig}
\usepackage{bm}

\usepackage{xmpmulti}

\def\pop{Phys.\ Plasmas\ }
\def\pre{Phys.\ Rev.\ E\ }
\def\prl{Phys.\ Rev.\ Lett.\ }
\def\pra{Phys.\ Rev.\ A\ }

\def\beq{\begin{equation}}
\def\eeq{\end{equation}}

\def\reff#1{(\ref{#1})}

\def\subsc#1{{\mbox{\rm\scriptsize #1}}}

\def\rhoc{\rho_\mathrm{c}}

\def\Wcmcm{\mbox{\rm Wcm$^{-2}$}}
\def\abl#1#2{\frac{\mbox{\rm d} #1}{\mbox{\rm d} #2}}
\def\abll#1#2{\frac{\mbox{\rm d}^2 #1}{\mbox{\rm d} #2^2}}

\def\N3d{N_\subsc{3D}}

\def\omegap{\omega_\mathrm{p}}

\def\vekt#1{\bm{#1}}
\def\vect#1{\vekt{#1}}
\def\vektr{\vekt{r}}

\def\vektp{\vekt{p}}
\def\vektv{\vekt{v}}

\def\pabl#1#2{\frac{\partial #1}{\partial #2}}

\def\atomu{\mbox{\rm \ a.u.\ }}

\def\Ea{E_\subsc{a}}

\def\Ta{T_\alpha}
\def\Tb{T_\beta}
\def\ma{m_\alpha}
\def\mb{m_\beta}
\def\mab{m_{\alpha\beta}}
\def\ea{e_\alpha}
\def\eb{e_\beta}
\def\va{v_\alpha}
\def\vb{v_\beta}

\def\na{n_\alpha}
\def\nb{n_\beta}
\def\lam{\ln{\Lambda}}

\def\vth{v_\mathrm{th}}

\def\ux{u_x}
\def\uy{u_y}
\def\uz{u_z}
\def\up{u_\perp}
\def\vp{v_\perp}
\def\sint{\sin\Theta}
\def\sinp{\sin\Phi}
\def\cost{\cos\Theta}
\def\cosp{\cos\Phi}
\def\nub{\nu_0}
\def\nus{\nu_\mathrm{s}}
\def\nud{\nu_\mathrm{d}}
\def\nut{\nu_\mathrm{t}}
\def\Ea{\mathcal{E}_\alpha}

\def\dtpic{{\Delta t}_\mathrm{PIC}}

\def\Ai{A\mathrm{i}}
\def\Bi{B\mathrm{i}}
\def\nuei{\nu_\mathrm{ei}}

\def\vth{v_\mathrm{th}}
\def\v0{v_0}
\def\vos{v_{os}}

\def\rhoc{\rho_\mathrm{c}}

\def\Te{T_\mathrm{e}}
\def\Ti{T_\mathrm{i}}

\def\Wcmcm{\mbox{\rm Wcm$^{-2}$}}
\def\abl#1#2{\frac{\mbox{\rm d} #1}{\mbox{\rm d} #2}}
\def\abll#1#2{\frac{\mbox{\rm d}^2 #1}{\mbox{\rm d} #2^2}}

\def\N3d{N_\subsc{3D}}

\def\omegap{\omega_\mathrm{p}}

\def\vekt#1{\bm{#1}}
\def\vect#1{\vekt{#1}}
\def\vektr{\vekt{r}}

\def\vektp{\vekt{p}}
\def\vektv{\vekt{v}}

\def\pabl#1#2{\frac{\partial #1}{\partial #2}}

\def\atomu{\mbox{\rm \ a.u.\ }}

\def\Ea{E_\subsc{a}}

\def\Ta{T_\alpha}
\def\Tb{T_\beta}
\def\ma{m_\alpha}
\def\mb{m_\beta}
\def\mab{m_{\alpha\beta}}
\def\ea{e_\alpha}
\def\eb{e_\beta}
\def\va{v_\alpha}
\def\vb{v_\beta}

\def\na{n_\alpha}
\def\nb{n_\beta}
\def\lam{\ln{\Lambda}}

\def\vth{v_\mathrm{th}}
\def\vos{v_\mathrm{os}}
\def\v0{v_0}
\def\I0{I_0}

\def\vg{v_\mathrm{g}}

\def\ux{u_x}
\def\uy{u_y}
\def\uz{u_z}
\def\up{u_\perp}
\def\vp{v_\perp}
\def\sint{\sin\Theta}
\def\sinp{\sin\Phi}
\def\cost{\cos\Theta}
\def\cosp{\cos\Phi}
\def\nub{\nu_0}
\def\nus{\nu_\mathrm{s}}
\def\nud{\nu_\mathrm{d}}
\def\nut{\nu_\mathrm{t}}
\def\Ea{\mathcal{E}_\alpha}

\def\dtpic{{\Delta t}_\mathrm{PIC}}

\def\Ai{A\mathrm{i}}
\def\Bi{B\mathrm{i}}
\def\nuei{\nu_\mathrm{ei}}
\def\nueibar{\overline{\nu}_\mathrm{ei}}



\usepackage{babel}

\begin{document}

\title{
Anomalous collisional absorption of laser light in plasma using particle-in-cell simulations
}
\author{M. Kundu}
\affiliation{Institute for Plasma Research, Bhat, Gandhinagar - 382 428, Gujarat, India }
\date{\today}
\begin{abstract}
Collisional absorption of laser light in a 
homogeneous, under-dense plasma is studied by a new 
particle-in-cell (PIC) simulation code considering one-dimensional 
slab-plasma geometry. Coulomb collisions between charge 
particles in plasma are modeled by a Monte Carlo scheme. 
For a given target thickness 
of a few times the wavelength of 800~nm laser of intensity $\I0$, 
fractional absorption ($\alpha$) of light due to Coulomb collisions 
(mainly between electrons and ions) is calculated at different 
electron temperature $\Te$ by introducing a total velocity 
$v = \sqrt{\vth^2 + \v0^2}$ dependent Coulomb logarithm 
$\ln\Lambda(v)$, where $\vth$, and
$\v0$ are thermal and ponderomotive velocity of an electron.
It is found that, in the low temperature regime 
($\Te\lesssim15$~eV), fractional absorption of light anomalously increases 
with increasing $I_0$ up to a maximum 
corresponding to an intensity $I_c$, and then it drops 
when $I_0>I_c$. 
Such an anomalous variation of $\alpha$ with $I_0$ in the low intensity regime 
was demonstrated earlier in experiments, and recently explained by 
classical and quantum models 
[Phys. Plasmas {\bf 21}, 013302 (2014); Phys. Rev. E {\bf 91}, 043102 (2015)].
Here, for the first time, we report anomalous collisional laser absorption by 
PIC simulations, 
thus bridging the gap between models, simulations, and experimental findings.

\end{abstract}
\maketitle

\section{introduction}\label{sec1}
One of the main objectives of the researchers working in the field of 
laser-plasma interaction (LPI) {\em is} 
to couple more laser energy with the plasma (or matter) so as to obtain
more energetic end-products, e.g., energetic charge particles or intense radiations. 
Therefore, it is of prime importance to know the underlying physical 
process (collisional and collisionless) by which laser energy is coupled 
to the plasma during the interaction. Earlier experiments 
\cite{bach,teubner,price,cerchez}
and theoretical studies \cite{rozmus,dong,kruerBook,mulserBook,shalomBook,gibbonBook}
have already reported various absorption processes, 
e.g., linear resonance \cite{manes77}, anharmonic resonance 
\cite{mulserx0,mulserx1,mulserx2,kundu2,kundu3,kundu4,kot03,cerchez}, 
Brunel heating \cite{brunel,mulserx3}, skin layer absorption \cite{bauer07}, 
$\vekt{J}\times \vekt{B}$ heating \cite{hong06} etc., which often depend on parameters of 
the laser, and the plasma 
\cite{gibbonBook,kruerBook,shalomBook,mulserBook}. 
For example, while passing through under-dense plasma 
(where plasma frequency $\omegap$ is less than the laser frequency 
$\omega$) an intense $p$-polarized short laser pulse can be absorbed 
by exciting wake-fields and instabilities
\cite{gibbonBook,kruerBook,shalomBook,mulserBook}. 
On the other hand, in an overdense plasma with an under-dense pedestal,
linear resonance absorption (LR) of $p$-polarized light may occur by 
meeting the resonance condition $\omegap=\omega$ in a 
specific location of the density gradient. 
Most often $p$-polarized light is used by experimentalists 
because of its ability to drive the plasma particles more efficiently, 
and relatively less attention is paid in LPI using $s$-polarized light. 
However, absorption of both $s$- and $p$- polarized light in plasma may happen
through the electron-ion collision \cite{bornath,bor01,kruerBook,shalomBook} 
known as inverse bremsstrahlung (IB) if laser intensity is below 
$10^{17}\,\Wcmcm$. 

In this work, we concentrate on the absorption of a 
$s$-polarized laser light in a homogeneous, under-dense plasma-slab due to IB, 
since collisional and collisionless absorption processes 
are coupled together for a $p$-polarized light where 
it is difficult to know what fraction of the collisional absorption 
contributes to the total absorption. We are also motivated by some earlier 
experimental results \cite{riley93,shalomBook} of collisional 
absorption with $s$-polarized light which shows that fractional absorption 
$\alpha$ of light (i.e., ratio of absorbed energy to the incident laser energy) 
anomalously increases initially with increasing laser intensity $\I0$ up to a 
maximum value about an intensity $I_c$, and then it drops nearly obeying the 
conventional scaling \cite{kruerBook,shalomBook}, i.e., $\alpha \propto I_0^{-3/2}$. 
Although there are numerous analytical models 
\cite{bor01,bornath,sch97,hil05,men13,mol12,pert72a,pert75,rand64,weng95,
silin,mulser1,mulser2,mulser3,kull01,wesson,pert95,rae92,catto77,
kremp01,brantov03,skupsky,riley93} which directly or indirectly describe 
conventional (standard) collisional absorption (CA) without the effect of background plasma, 
less attempts were made to examine above mentioned anomalous collisional 
absorption (ACA).
Recently, ACA process has been explained in the low temperature regime ($\Te\lesssim15$~eV) 
by postulating a total velocity dependent Coulomb logarithm $\ln\Lambda(v)$ 
(where $v = \sqrt{\vth^2 + \v0^2}$, $\vth$ 
and $v_0$ are the thermal and ponderomotive velocity) in
an analytical model of electron-ion collision frequency $\nuei$ \cite{mulser1,kundu1} and more rigorous kinetic treatment \cite{kunduPRE}.
However, to what extent this analytical approach is valid can {\em only} be answered
numerically with self-consistent dynamics of plasma background under the 
laser irradiation.
To this end, we have developed an one-dimensional electromagnetic particle-in-cell 
code (henceforth we call 
it EMPIC1D) where variation of physical quantities (charge density, current density, 
electro-magnetic fields) depend only on the one spatial coordinate along the laser
propagation direction while considering all three velocity components of charge
particles. 
%
In a particle-in-cell (PIC) simulation a reduced number of computational particles is used 
to represent plasma, instead 
of a large number of actual physical particles \cite{birdsall,hockney}. 
This technique reduces the computational load, and enables to 
study the dynamics of an actual physical system of large number of
charge particles. Sizes of these 
PIC particles (computational particles) are typically on the order of a 
numerical grid, they can pass through each other during the 
interaction, and Coulomb collisions do not naturally 
happen \cite{sentoku,cadjan,takizuka,takizuka2,sma,manheimer,oliphant}.
For this reason, Coulomb collisions are explicitly added in all PIC codes.
To include Coulomb collision in our EMPIC1D code a Monte Carlo (MC) 
technique proposed by Takizuka and Abe \cite{takizuka} is adopted. Recently, this scheme is used in the PARASOL electrostatic PIC code \cite{takizuka2} to study kinetic effects in tokamak plasmas. It conserves total energy 
and total linear momentum before and after a collision event in the velocity space. 

In this work, with the Monte Carlo collision assisted EMPIC1D code, 
for the first time we show the aforementioned anomalous absorption in an 
under-dense plasma in the low temperature regime similar to our earlier 
analytical works \cite{kundu1,kunduPRE}. 
Thus we bridge the gap between the experimental findings, 
analytical models, and PIC simulations.

Plasma is assumed to be pre-ionized. 
Laser intensity is kept below $10^{18}\,\Wcmcm$ so that relativistic 
effects are less important.
For convenience, atomic units (a.u.) are used unless mentioned explicitly, 
i.e., $-e = m = 4\pi\epsilon_0 = \hbar = 1$, where $\vert e \vert$ is the
electronic charge and mass, $\epsilon_0$ is the permittivity of free space, 
and $\hbar$ is the reduced Planck constant.

This article is organized in the following manner. Details of the EMPIC1D code 
is given in Sec.\ref{sec2}, with appropriate benchmarking 
in the absence of collision. In Sec.\ref{sec3} we independently benchmark the 
Monte Carlo (MC) binary collision scheme. Then we combine the collision module with 
the EMPIC1D code, and study collisional absorption of a $s$-polarized light 
in an under-dense plasma slab in Sec.\ref{sec4} where a comparison 
is also made between simulation and theoretical results. 
A summary is given in Sec.\ref{sec5}.

\section{Details of the PIC code}\label{sec2}
\begin{figure}
\includegraphics[width=0.45\textwidth]{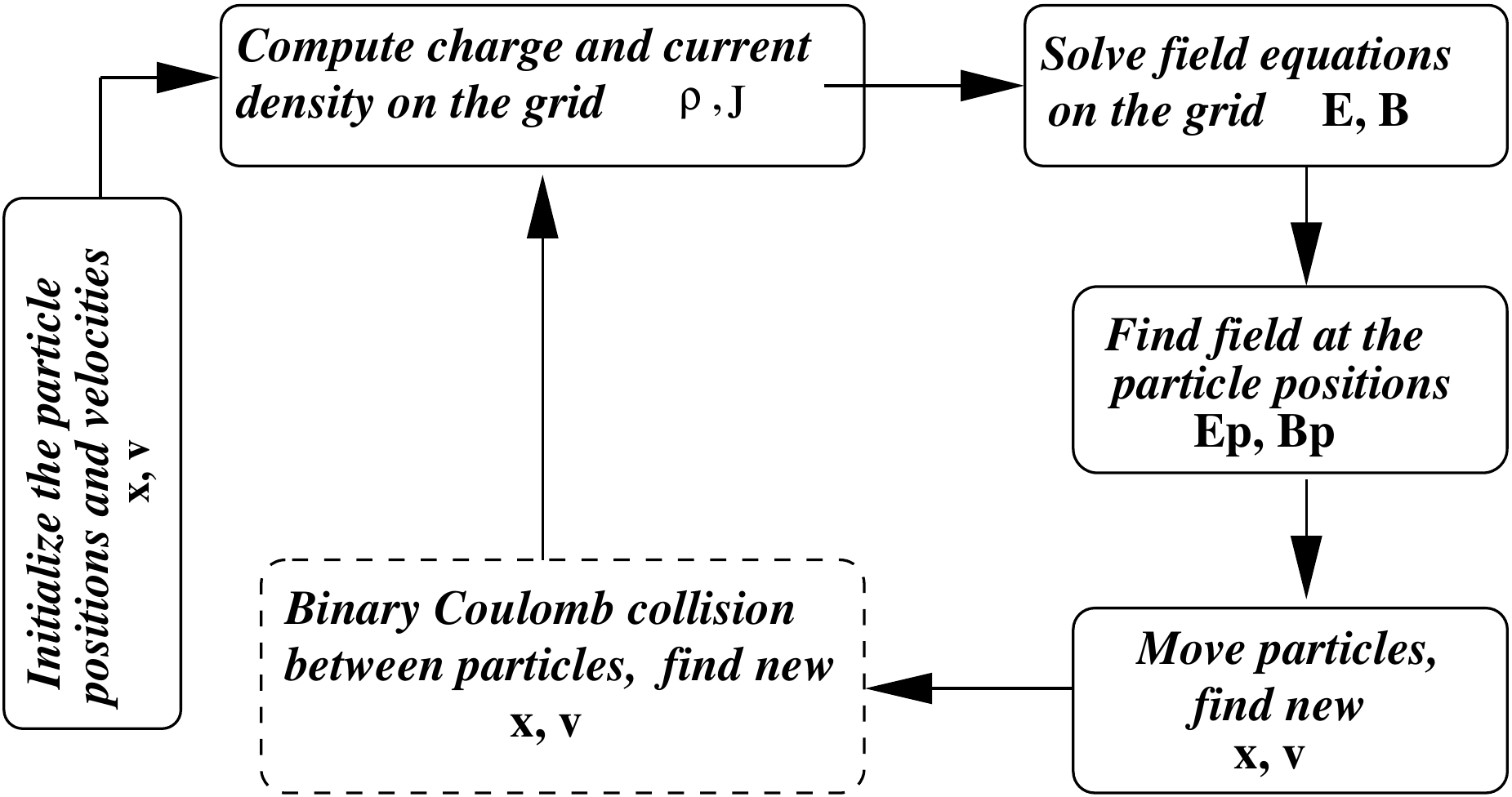}
\caption{Schematic of a PIC simulation with binary collision.}
\label{picCycle}
\end{figure}

%
Here we give only necessary details of the PIC code. 
In PIC
a collection of physical particles is represented by a 
computational particle so that the charge to mass ratio $q/m$ of the 
computational particle remains same as that of a physical particle. 
The following Maxwell-Lorentz system of equations (in the normalized form) 
is solved numerically after the discretization in space and time :
\begin{eqnarray}
\label{eq1}
\pabl{\bar {\vect B}}{t} &=& -c{\vect\nabla}\times {\vect E}, \\
\label{eq2}
\pabl{\vect E}{t} &=& c{\vect\nabla}\times {\bar {\vect B}} - 4\pi {\vect J},\\
\label{eq3}
\dot{\vektp} &=& q\left({\vect E_p} 
+ {\vect v}\times{\frac{\bar{\vect B}_p}{c}}\right). 
\end{eqnarray}
Here, ${\vekt E}, {\bar{\vect B}}$ are the electric and magnetic part of the electromagnetic field, $\vektp$ is the particle momentum corresponding to 
its velocity $\vektv$ and position $\vektr$ at a time $t$. ${\vect J}$ is the
current density vector, $c$ is the speed of light in the free space. The 
scaling ${\bar {\vect B}} = c{\vect B} $ connects actual magnetic 
field ${\vect B}$ with the scaled magnetic field ${\bar{\vect B}}$. 
The other equations, namely,
${\vect\nabla}\cdot {\bar {\vect B}} = 0$ and the Gauss's law
${\vect\nabla}\cdot {{\vect E}} = 4\pi\rho$ are not explicitly solved
in a standard multi-dimensional PIC scheme, thus saving a substantial amount of computer time. However, ${\vect\nabla}\cdot {\bar {\vect B}} = 0$ is ensured by choosing a staggered grid, called Yee mesh. 
The charge and current conservation follows from 
$\partial{\rho}/\partial{t} + {\vect\nabla}\cdot {\vect J} = 0$, thus ensuring 
${\vect\nabla}\cdot {{\vect E}} = 4\pi\rho$. 
Note that ${\vekt E},{\bar {\vekt B}}, {\vekt J}, \rho$
are calculated on the grid points. 
Therefore ${\vekt E}, {\bar{\vekt B}}$ field are interpolated 
to obtain corresponding fields ${\vekt E_p}, {\bar{\vekt B}}_p$ at the particle 
(particle charge $q$ and mass $m$) position ${\vect r}$
using linear weighting scheme, and the Lorentz equation \reff{eq3}
is solved using the standard leap-frog method. 
The advantage of the scaling ${\bar {\vect B}} = c{\vect B} $ is that, it reduces 
equations \reff{eq1} and \reff{eq2} identical in form in the free 
space (i.e., when $\vect{J} = 0$) and the amplitudes of $\vect{E}, \vect{\bar{B}}$ becomes comparable. 
From now onward, for convenience, we shall write $\vect{B}$ 
in stead of $\vect{\bar {B}}$ unless mentioned explicitly.
A typical cycle of the PIC simulation is shown in Fig.\ref{picCycle} 
which clearly depicts where the binary collision module (the MC part)
should be incorporated.

\subsection{Simplification in one-dimension}\label{sec2a}
Let us consider a {\em s}-polarized light (propagating in $y$-direction) 
with transverse field components $E_z, B_x$. The physical quantities 
(e.g., charge density, current density, electro-magnetic fields)
are assumed to depend only on the space co-ordinate $y$, while retaining all three 
velocity components ($v_x,v_y,v_z$) of particles.
Components of Eqn.\reff{eq1}-\reff{eq2} reads
\begin{eqnarray}
\label{eq4}
\pabl{{B_x}}{t} &=& -c \pabl{E_z}{y}, \\
\label{eq5}
\pabl{E_z}{t} &=& -c \pabl{{B_x}}{y} - 4\pi {J_z(t,y)},\\
\label{eq5b}
\pabl{E_y}{t} &=& - 4\pi {J_y(t,y)}.
\label{eq6l}
\end{eqnarray}
Equation \reff{eq6l} gives longitudinal component of the electric field
$E_y$ in our case. 
It is important to mention that, the numerical implementation of our PIC code 
is little different from some of the traditional 1D-PIC codes, namely, EM1BND \cite{birdsall}, 
LPIC++ \cite{gibbonBook,lichters}, but closely follow the implementation 
in PIC codes PSC \cite{harmut}, VLPL \cite{pukhov}, VPIC \cite{bowersVPIC}, OSIRIS \cite{foncecaOSIRIS},
and VORPAL \cite{nieterVORPAL}. In EM1BND \cite{birdsall} and LPIC++ \cite{gibbonBook,lichters},
by performing addition and subtraction of Eqn.\reff{eq4} and \reff{eq5}, and 
writing $\psi_{\pm} = E_z\pm B_x$ one finds 
$(\partial_t \pm c\partial_y)\psi_{\pm} = - 4\pi J_z(t,y)$; where $\psi_{\pm}$ 
can be recognized as the two 
propagating solutions of the wave equation. The advantage in this traditional 
procedure \cite{birdsall,gibbonBook,lichters} is that the partial derivative 
$(\partial_t \pm c\partial_y)$ can be written in terms of the total derivative in time 
w.r.t. an observer moving at a speed $\pm c$, leading to  
\begin{equation}
\label{eq5a}
\abl{\psi_{\pm}}{t} = - 4\pi J_z(t,y).
\end{equation}
For a given $J_z$, Eq.\reff{eq5a} is solved as an ordinary differential 
equation (ODE) to obtain transverse fields 
$E_z =(\psi_{+} + \psi_{-})/2$, $B_x=(\psi_{+}-\psi_{-})/2$ on the grid.
It also allows larger time step $\Delta t = \Delta y /c$.
The disadvantage is that, the longitudinal component $E_y$ is 
obtained by solving the Poisson's equation 
${\partial E_y}/{\partial y} = 4\pi\rho$ explicitly (not from Eq.\reff{eq6l}), and 
it needs a separate algorithm than solving the transverse components. 
Moreover, this traditional scheme is hard to extend in multi-dimensional case.
In our EMPIC1D code we use FDTD (finite difference in time domain) method
for the solution of all field components (both traverse and longitudinal), 
in stead of the traditional addition-subtraction method mentioned above. 
Thus we use only one kind of algorithm for $E_z, B_x, E_y$ which is extendable
to PIC simulations in higher dimensions as in Refs.\cite{harmut,bowersVPIC,pukhov,foncecaOSIRIS,nieterVORPAL}.
Using FDTD procedure on the Yee mesh, and assuming 
$t=n\Delta t$, $y=k\Delta y$, 
Eqn.\reff{eq4},\reff{eq5},\reff{eq6l} can be written as 
\begin{widetext}
\begin{eqnarray}
\frac{{B_x^{n+1}(k+1/2)} - {B_x^{n}}(k+1/2)}{\Delta t} & = & \label{fdtdB} 
-c \frac{E_z^{n+1/2}(k+1) - E_z^{n+1/2}(k)}{\Delta y},\\ 
\frac{E_z^{n+1/2}(k+1) - E_z^{n-1/2}(k+1)}{\Delta t} & = & \label{fdtdE}
-c \frac{{B_x^{n+1}(k+1/2)} - {B_x^{n+1}}(k-1/2)}{\Delta y} 
- 4\pi {J_z^{n+1/2}}(k+1),\\ 
\label{fdtdEy}
\frac{E_y^{n+1/2}(k+1) - E_y^{n-1/2}(k+1)}{\Delta t} & = & 
- 4\pi {J_y^{n+1/2}}(k+1). 
\end{eqnarray}
\end{widetext}

To ensure numerical stability we take $c\Delta t/\Delta y = 1/2$, which decides
the time step $\Delta t$ for a chosen grid size $\Delta y$. The dispersion due to the 
FDTD discretization is minimized by choosing sufficient number of spatial grids 
(minimum 40 is taken) per wavelength of light. VLPL like \cite{pukhov} 
improvement, namely, dispersion free scheme we plan to include in future.
The current density $\vect{J}$ due to the 
motion of charge particles is computed using the 
``explicit current conserving scheme'' by Umeda et al. \cite{umeda} which 
satisfies $\partial{\rho}/\partial{t} + {\vect\nabla}\cdot {\vect J} = 0$. 
Thus we avoid explicit solution of the Poisson's equation to obtain $E_y$. 
We use ``perfectly matched layer'' (PML) absorbing boundary condition 
\cite{sullivan} for the electromagnetic fields. For charge particles, however, 
depending upon physical situations, absorbing, periodic, and 
reflecting boundary conditions are used.  

\subsection{Benchmarking of the PIC code}\label{sec2b}
The EMPIC1D code is verified for different cases (where analytical solutions exist) : 
(i) plasma oscillation and energy conservation (without external light source), 
(ii) reflection of the laser light interacting with an over-dense plasma and 
corresponding field in the skin layer, 
(iii) transmission and reflection coefficient of laser light in an 
under-dense plasma, and (iv) interaction of light with inhomogeneous plasma 
and formation of standing waves. 
As a representative case, for the purpose of benchmarking, 
here we consider only the last case. 

\subsubsection{Interaction of s-polarized light with inhomogeneous plasma}
\begin{figure}
\includegraphics[width=0.235\textwidth,height=1.6in]{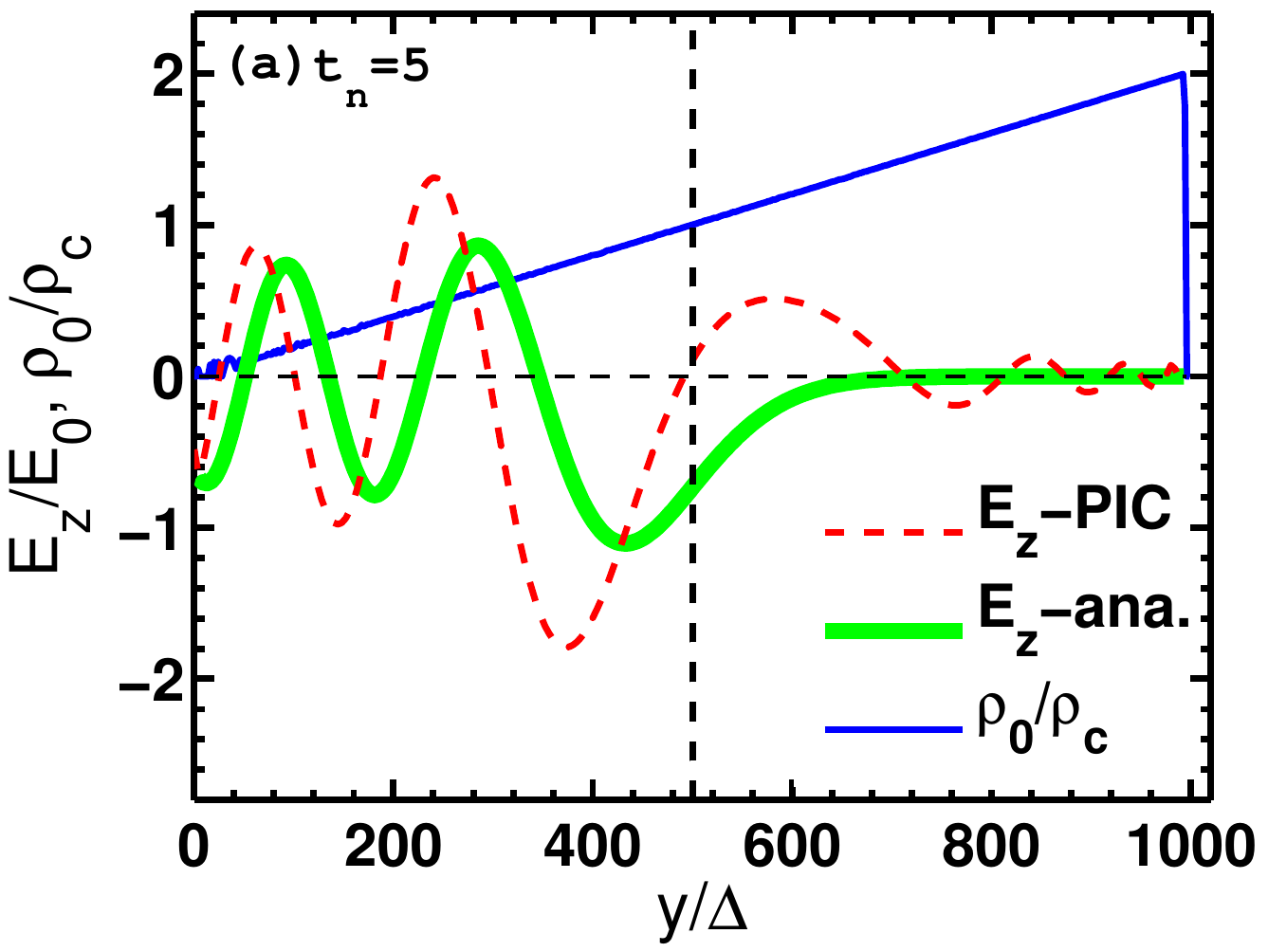}
\includegraphics[width=0.235\textwidth,height=1.6in]{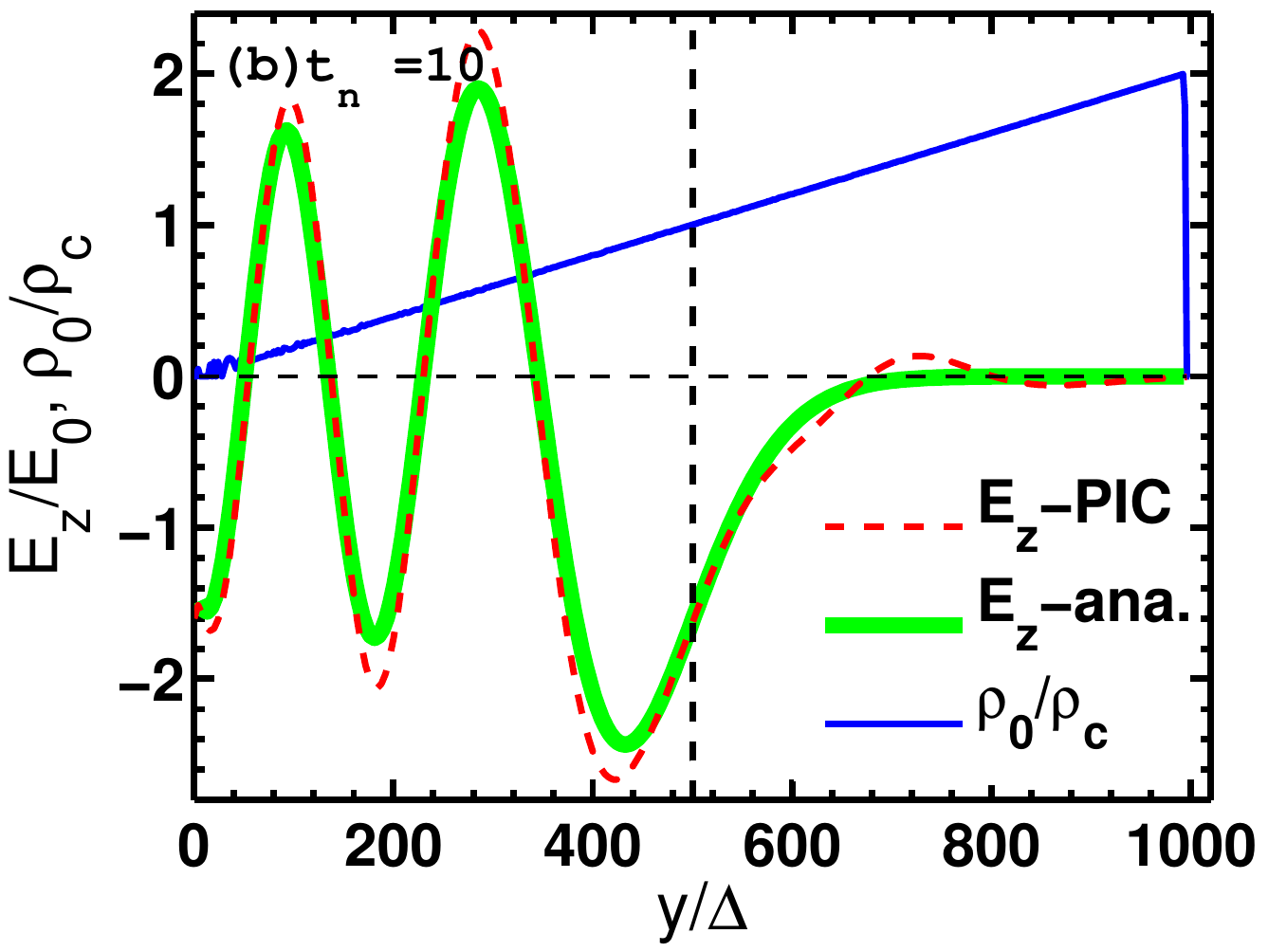}
\includegraphics[width=0.235\textwidth,height=1.6in]{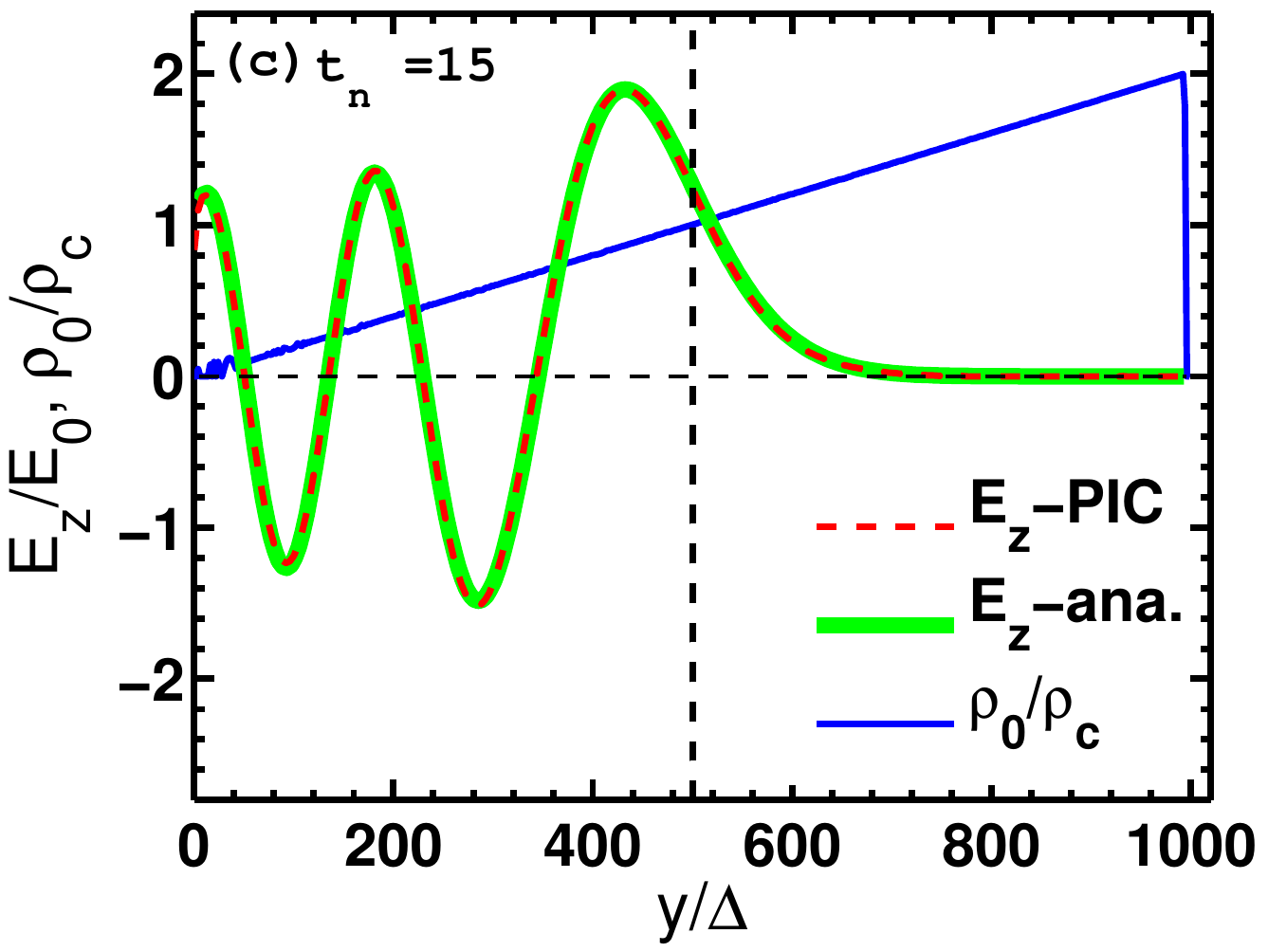}
\includegraphics[width=0.235\textwidth,height=1.6in]{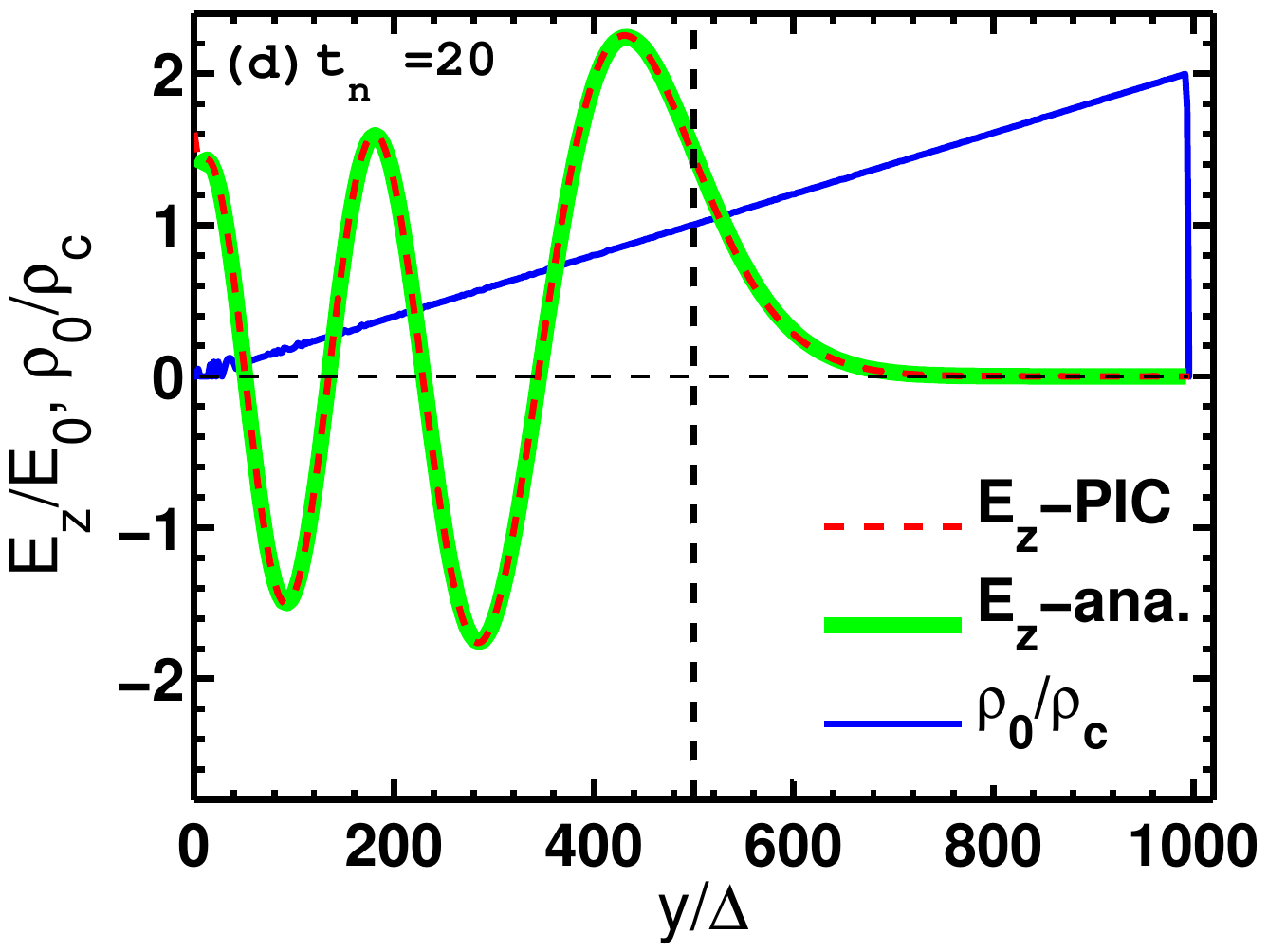}
\caption{(Color online) Formation of standing wave solution due to the 
interaction of s-polarized light with inhomogeneous plasma of 
linear density profile $\rho_0/\rhoc$ (dark solid line, blue).
Vertical dashed line is the critical density layer.
At an early time $t_n=5$ in (a) the profile of the
laser field $E_z$ in the PIC simulation (dashed oscillatory line, red) 
is far from the steady state Airy profile (solid gray line, green). 
As time increases difference between PIC profile and the analytical 
profile gradually decreases which is clear from (a) and (b).
After a long time, e.g., $t_n=15, 20$ in (c,d) 
PIC result matches the desired analytical Airy profile.}
\label{airyfig}
\end{figure}

Consider the interaction of 
a laser light with an inhomogeneous plasma which consists of an under-dense 
region and an over-dense region separated by a critical density ($\rho_c = \omega^2/4\pi$) surface,
illustrated in Fig.\ref{airyfig}. This problem was investigated analytically 
in Refs.\cite{kruerBook,shalomBook,Ginzburg} for time independent case. 
However its numerical verification with PIC method
is still rare in the literature. Benchmarking of this problem 
{\em alone} can justify the correctness of interaction 
of laser field with under-dense and over-dense plasma in a single 
simulation run using EMPIC1D.   
We assume that density variation of plasma is linear
$\rho_0(y) = \rhoc y/L$ along the propagation direction $y$ 
of a $s$-polarized light described by $E_z,B_x$. 
$L$ is the distance of the critical density surface (vertical dashed line)
from the vacuum plasma interface on the left side. 
A continuous laser light $E_0\sin(\omega t)$ 
of amplitude $E_0$ is sourced through the vacuum
plasma interface. After propagating through the under-dense region $\rho_0(y)/\rho_c<1$, light is 
reflected from the critical surface. The reflected light and the incident 
light are superimposed to form a standing wave after a 
long time when steady state is reached, and 
the resultant field $E_z$ satisfies the wave equation \cite{kruerBook,shalomBook}
$\abll{E_z}{\eta} - \eta E_z = 0$ 
with $\eta(y) = (\omega^2/c^2 L)^{1/3}(y-L) $. 
Analytical solution $E_z(\eta) = c_1 \Ai(\eta) + c_2 \Bi(\eta)$ 
is found in terms of 
Airy functions $\Ai,\Bi$ with $c_1,c_2$ as constants (independent of $\eta$). 
For the standing wave solution $c_2 = 0$ is chosen, since
$\Bi\rightarrow\infty$ as $\eta\rightarrow\infty$, leading to
$E_z(\eta) = c_1 \Ai(\eta) $. In reality, the problem is time dependent. Desired steady state analytical solution may not be reached during the early 
time of interaction. The constant $c_1$ depends upon the 
amplitude and phase of the resultant wave at the vacuum plasma interface at a given instant of time. 
We adjust $c_1$ with
the effective magnitude of the field $E_s^{t_n}$ at a time $t=t_n$ 
retrieved from the PIC simulation at the vacuum plasma interface (at $y_l$) 
to match the analytical solution with the PIC simulation such that 
$c_1(t_n)=E_s^{t_n}(y_l)/\Ai(\eta(y_l))$. It leads to the analytical solution as
$E_z^{t_n}(y)= \left[E_s^{t_n}(y_l)/\Ai(\eta(y_l))\right]\Ai(\eta(y))$.  

We take laser field parameters as $E_0=0.2$\atomu, and $\omega=0.057$\atomu corresponding 
to $\I0 \approx 7 \times 10^{15}\,\Wcmcm$ and $\lambda=800$~nm respectively.
The length of the simulation box is $L_b=1000\Delta$ ($\Delta = \Delta y = \lambda/40$ 
is the grid size) with total number of 
grids $N_g=1000$. Out of these 1000 grids, plasma particles centrally occupy 
990 grids while 5 grids are left on each boundaries to separate the plasma 
from the vacuum. The source is located at the left plasma vacuum interface at
$y=y_l = 5\Delta$. The right boundary of the plasma is at $y=y_r = 995\Delta$.
There are $N=7920$ PIC ions (assumed to be stationary)
each of charge $q_s= 0.00325$\atomu and an equal number of overlapping PIC electrons 
with mass $m_s = q_s$, and charge $-q_s$. 
Particles are loaded according to the scheme given in Ref.\cite{birdsall} 
to obtain a linear density profile as shown in Figs.\ref{airyfig}(a)-(d)
with $\rho_0/\rhoc$ rising linearly from 0 to 2 (dark solid line, blue). The field 
profiles $E_z/E_0$ from PIC simulation (dashed oscillatory, red) and the 
analytical solution (solid oscillatory, green) are plotted 
against $y/\Delta$ in Figs.\ref{airyfig}(a-d) 
at different times $t_n=5,10,15,20$ respectively.
At an early time $t_n=5$, PIC solution for $E_z$ does not match with the
analytical solution (in Fig.\ref{airyfig}a) since at this early time superposition of the 
reflected wave with the incident wave is still incomplete in the 
PIC simulation to form the desired standing wave. 
As time increases, difference between the PIC profile and the analytical 
profile gradually decreases which is evident from the comparison between 
Fig.\ref{airyfig}a and Fig.\ref{airyfig}b.
After a sufficiently longer
time $t_n=15$ and beyond, the reflected wave meets 
the incident wave with required amplitude and phase so that the PIC 
solution matches (in Figs.\ref{airyfig}c,d) the analytical solution (see 
oscillatory dashed line coincides the solid line). 
The match between analytical and the numerical solution clearly ensures 
the correctness of our PIC simulation without collision.
\section{Binary collision: model and simulation}\label{sec3}
\begin{figure}
\includegraphics[width=0.3\textwidth]{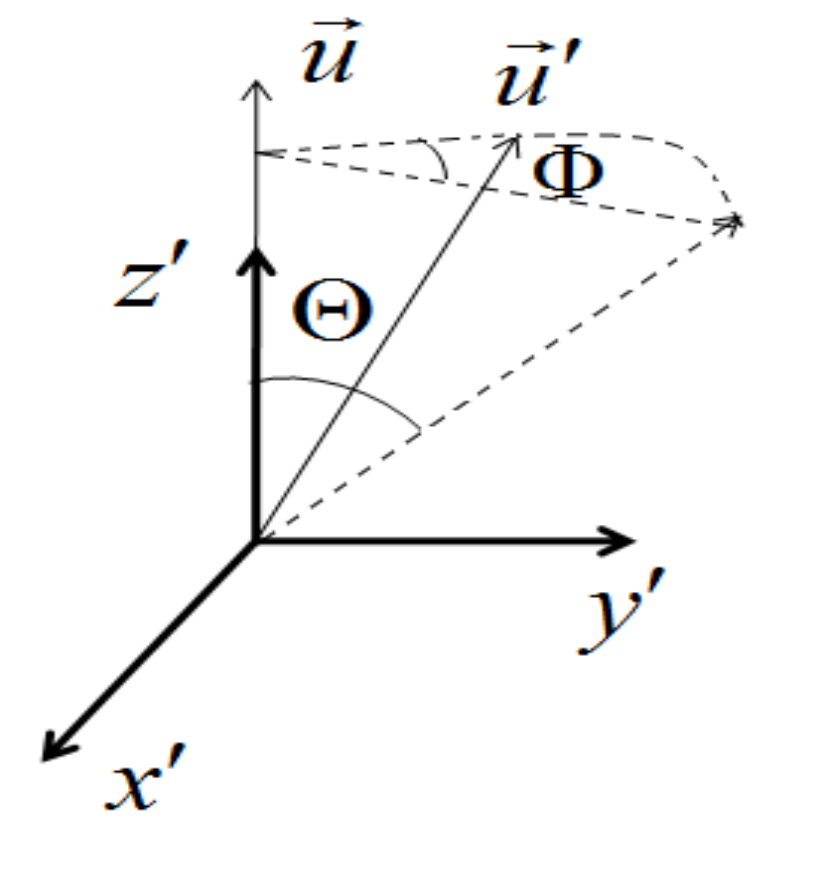}
\caption{Relative velocities $(\vec{u},\vec{u}')$ before and after the collision.}
\label{velRot}
\end{figure}

In the PIC simulation, as discussed above, particles may pass through each other during 
their close encounter and collision effects are omitted. 
To implement binary collision in the EMPIC1D code we have followed the 
Monte Carlo scheme given by Takizuka and Abe \cite{takizuka} and also the work 
by Ma et al. \cite{sma}, and Sentoku et al. \cite{sentoku}. 
For conciseness we only show the validation of our implementation with a minimum detail. 
The main approximation of binary collision is that at a given instant only 
two particles will collide, and the effect of collision arises due to the 
cumulative effect of many small angle binary collisions. Within a computational 
cell, particles are paired randomly (ion-ion, ion-electron, electron-electron) 
and then collision is performed between every pair. The maximum impact parameter 
in a fully ionized quasi-neutral plasma being of the order of the Debye length 
\cite{truvnikov}, 
the maximum size of the collision grid is also restricted to the Debye length. 
Collision event takes place in the velocity space, meaning that 
the velocity components of the particles changes but the co-ordinates 
are not influenced at that time. The post collision velocities are obtained 
by going to the center of mass (COM) frame of the respective collision pairs 
and then back to the laboratory frame.
Due to collision, during a small time interval $\Delta t_c$ (which is 
sufficiently small compared to the mean relaxation time), the direction of
velocities of the colliding particles changes but not their magnitudes. 
For instance, we consider a system of two particles from two species 
$\alpha$ and $\beta$ having velocities $\va$ and $\vb$, masses $\ma$ 
and $\mb$, densities $\na$ and $\nb$, and charges $\ea$ and $\eb$. 
At a given instant $t$, the effect of collision leads to the rotation of 
the relative velocity ${\vect u} = {\vect v}_\alpha - {\vect v}_\beta$ 
in the COM frame of the two particles. 
The relative velocity ${\vect u'} = {\vect v'}_\alpha - {\vect v'}_\beta$ 
after the collision (primes represents quantities after collisions) in the COM 
frame, and the rotation of the velocity vector ${\vect u}\rightarrow {\vect u'}$
can be described by the scattering angle $\Theta$ and the azimuth angle 
$\Phi$ (see Fig.\ref{velRot}) which are chosen randomly for a given pair ($\alpha,\beta$). 
In order to find $\Theta$, a parameter $\delta$ is introduced 
such that $\delta = \tan(\Theta/2)$.
The variable $\delta$ is chosen randomly from a Gaussian distribution
such that its mean is zero, and 
corresponding variance $\left<\delta^2\right>$ is \cite{takizuka,takizuka2,sma}
\beq
\left<\delta^2\right> = \frac{\ea^2\eb^2 n_L\lam}{8 \pi\epsilon_0^2\mab^2 u^3}\Delta t_c.
\eeq
Here $n_L$ is the minimum density between $n_\alpha$ and $n_\beta$,  
$\mab=\ma\mb/(\ma + \mb)$ is the reduced mass, and $\ln \Lambda$ is the 
Coulomb logarithm.  
$\ln \Lambda$ can be calculated 
as \cite{truvnikov}
\beq
\lam = \ln[{\lambda_D(\Ta + \Tb)}/{2\ea\eb}] 
\label{Clg}
\eeq
where $\lambda_D$ is the Debye length, and $T_\alpha, T_\beta$ are the temperatures of the respective species. Above expression of $\lam$ is valid 
in the absence of external laser field. Otherwise,
$\ln \Lambda$ should include the response of electrons to the 
laser field strength $E_0$ and the frequency $\omega$. We shall use Eq.\reff{Clg} 
only for the validation of binary collision event when there is no external force. 
The necessary modification of $\lam$ with laser field will be discussed later. 
Deflection angle $\Theta$ is calculated by using Box-Muller method 
with distribution 
$p(\delta) d\delta = (1/\langle\delta^2\rangle) \exp(-\delta^2/2\langle\delta^2\rangle) \delta d\delta$ as given in Refs.\cite{pert99,cohen2013},
%
%
\beq
\label{dangle}
\Theta = 2\arctan\sqrt{-2\left<\delta^2\right>\ln(1-R_1) },
\eeq
where $R_1$ is an uniform random number between 0 and 1.
The azimuth angle $\Phi$ is chosen as $\Phi= 2\pi R_2$, with $R_2$ 
as an uniform random number between 0 and 1. 
The change in velocity components in the 
laboratory frame can be calculated as \cite{takizuka}
%
\begin{eqnarray} 
\nonumber
\Delta \ux &=& \frac{\ux}{\up}\uz\sint\cosp 
- \frac{\uy}{\up}u\sint\sinp \\ 
& - & \ux(1-\cost)\\
\nonumber 
\Delta \uy &=& \frac{\uy}{\up}\uz\sint\cosp 
+ \frac{\ux}{\up}u\sint\sinp \\ 
& - & \uy(1-\cost) \\
\Delta \uz &=& -\up\sint\cosp -\uz(1-\cost) 
\end{eqnarray}
where $\up = \sqrt{\ux^2 + \uy^2}$. When $\up = 0$, we take 
\begin{eqnarray} 
\Delta \ux &=& u\sint\cosp, \\
\Delta \uy &=& u\sint\sinp, \\
\Delta \uz &=& -u(1-\cost). 
\end{eqnarray}
Final post collision velocities in the laboratory frame reads 
\begin{eqnarray} 
\label{velcol1}
\vect{v}_{\alpha}({t+\Delta t_c}) &=& \vect{v}_{\alpha}(t) + (\mab/\ma)\Delta \vect{u} \\ 
\label{velcol2}
\vect{v}_{\beta}({t+\Delta t_c}) &=& \vect{v}_{\beta}(t) - (\mab/\mb)\Delta \vect{u}. 
\end{eqnarray}

\subsection{Validation of the binary collision model}
\begin{figure}
\includegraphics[width=0.45\textwidth]{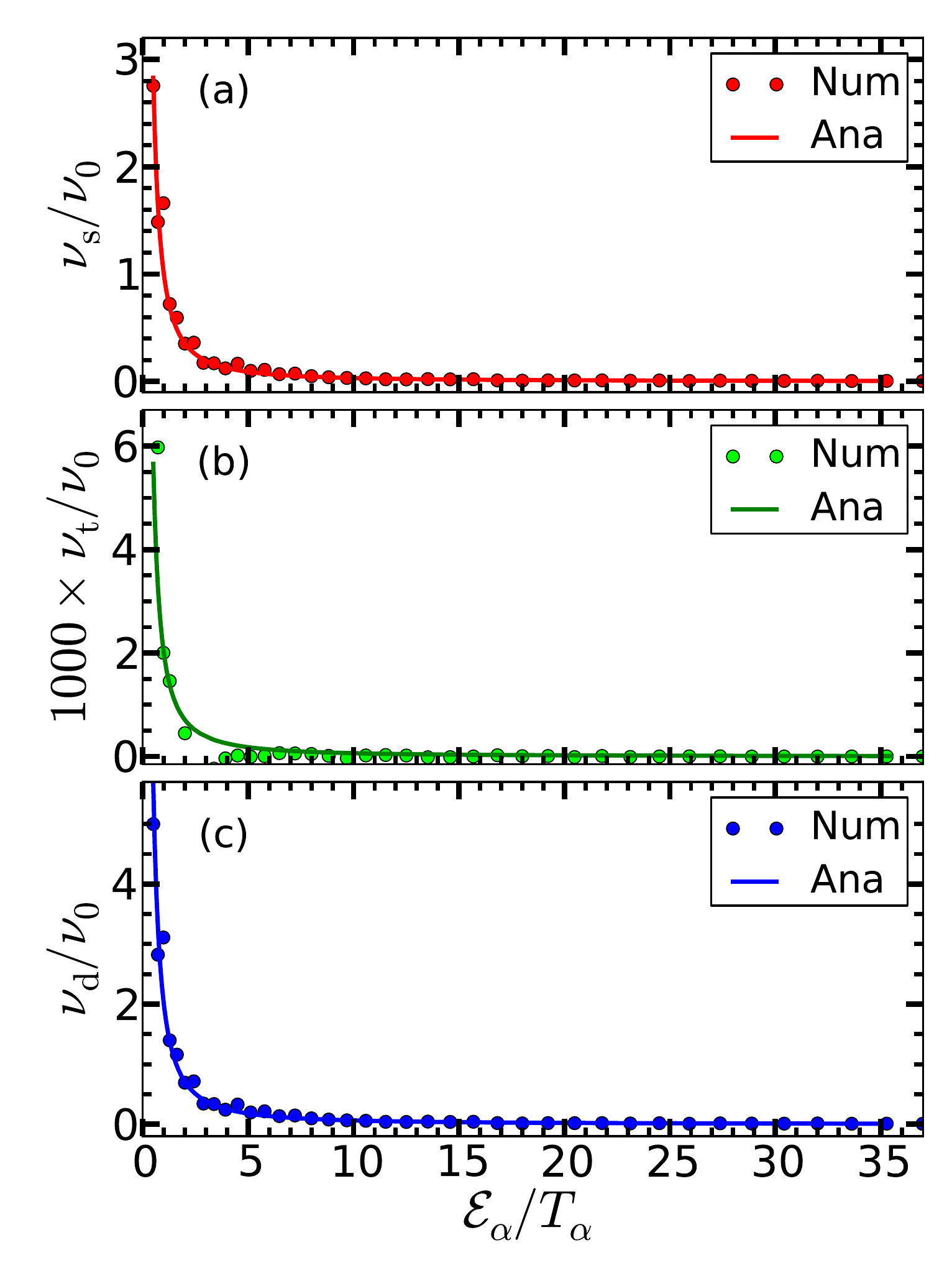}
\caption{Normalized (a) slowing down frequency $\nu_s$, (b) energy transfer frequency $\nu_t$, 
and (c) deflection frequency $\nu_d$ vs normalized
energy $\mathcal {E}_\alpha/T_\alpha$ of electrons due to Coulomb collision with 
background ions after $n_t = 20$ time steps. Each numerical point represents one group of 
particles (out of total 40 groups) each having 250 particles. 
Solid lines are analytical results using Eqns.\reff{nuseq1}-\reff{nudeq1} while circles represent respective numerical results.}
\label{fig3}
\end{figure}

In order to validate our implementation of binary collision, we consider particle collisions 
in a two component plasma. The conservation of energy, and momentum for each colliding pair 
of particles are rigorously checked. We benchmark different relaxation rates 
(collision frequencies): 
(i) mean rate of change of velocity of the electrons (called slowing down frequency $\nus$), 
(ii) mean rate of change of energy of the test electrons (energy transfer frequency, $\nut$), and 
(iii) the rate of spread of the test particle velocity transverse to its 
original direction (deflection frequency, $\nud$). 
Analytically, these frequencies are obtained by the test particle theory 
\cite{truvnikov,spitzer} in which a test particle (electron, designated by 
$\alpha$) is assumed to move through a medium of field particles 
(ions, designated by $\beta$) having Maxwellian velocity distribution.
These frequencies are related to the transport equations \cite{truvnikov,spitzer}
\begin{eqnarray} 
\label{nuseq}
\left<\abl{\va}{t}\right> &=& -\nus\va, \\
\label{nuteq}
\left<\abl{\Ea}{t}\right> &=& -\nut\Ea, \\
\label{nudeq}
\left<\abl{\vp^2}{t}\right> &=& \nud\va^2,
\end{eqnarray}
and can be expressed in terms of the integral  
$\mu(x) = \frac{2}{\pi}\int_0^x \sqrt{t}\exp(-t)dt$ and 
its derivative $\mu'(x) = \frac{2}{\pi} \sqrt{x}\exp(-x)$ as
\begin{eqnarray} 
\label{nuseq1}
\nus &=& \nub\left(1 + \ma/\mb\right)\mu(x) \\
\label{nuteq1}
\nut &=& 2\nub\left(\ma\mu(x)/\mb - \mu'(x)\right) \\
\nud &=& 2\nub\left(\mu(x) + \mu'(x) - \mu(x)/{2 x} \right). 
\label{nudeq1}
\end{eqnarray}
Here $\nub = {\ea^2\eb^2\nb\lam}/\left({8\pi\epsilon_0^2\sqrt{2\ma\Ea^3}}\right)$ 
is the basic collision frequency, $\Ea = \ma\va^2/2$ is the kinetic energy of the 
projectile, and $x=\mb\Ea/{\ma\Tb}$. To obtain $\nus, \nut, \nud$ an average over 
a group of test particles (electrons) having the same velocity are taken
for the better statistics. 

We consider 10000 test particles (electrons) and an equal number of
field particles (ions). Test particles are assumed to be 
composed of 40 groups, and each of 250 particles in a 
group has the same initial velocity. Their velocities are normalized by
the thermal velocity, $\vth = \sqrt{\Ta/\ma}$. 
Here we assume $\mb/\ma=10000$ (i.e. the field particles are highly massive), 
$\Ta=\Tb=T=5$~eV, $\ea=\eb=e$ (charge of the electron), $\na=\nb=n_d=10^{15}\,\mathrm{cm}^{-3}$, 
and $\nub\Delta t_c = 10^{-3}$. 
The value of $\nub\Delta t_c$ is chosen small to ensure small angle collisions \cite{sma}. 
In the Monte Carlo simulation, $\nus,\nut,\nud$ are calculated using 
the Eqn.\reff{nuseq},\reff{nuteq}, and \reff{nudeq} with the velocities 
from Eqn.\reff{velcol1}, and \reff{velcol2} before and after collisions.
%

Figure~\ref{fig3} shows the normalized frequencies ($\nus/\nub, \nut/\nub$,  
and $\nud/\nub$) versus the energy of 
the test particles $\mathcal{E}_\alpha/T_\alpha$ after $n_t=20$ time steps. 
Each solid circle (numerical) represents the average over a group of test particles. 
An excellent agreement between the numerical results
and the analytical values (solid lines) 
from Eqn.\reff{nuseq1},\reff{nuteq1}, and \reff{nudeq1} ensures the 
correctness of our implementation. 

\section{Absorption in an under-dense plasma}\label{sec4}
We have integrated the above collision module with the EMPIC1D
code described in Sec.\ref{sec2} to study 
collisional absorption of light 
incident normally on an under-dense plasma slab of uniform density. 
The simulation domain consists of $N_g = 500$ computational cells with
the plasma slab 
at the center. 
Initially each computational cell contains 
equal number of electrons and ions so that plasma 
is charge neutral. The temporal profile of the laser pulse 
(at the left boundary, $y_l$) is chosen as 
\beq
E_z(t,y_l) = E_0 
\left\{\begin{array}{ll}
\sin^2(\omega t/2 n_c)\cos(\omega t);\;\; & 0<t<n_c T \\
 0\;\; & t>n_c T, \\
\end{array}\right. \label{eqSin2}
\eeq
with $n_c$ as the number of cycles, and $T = 2\pi/\omega$ as the laser period.  
The pulse is numerically excited at $y=y_l$, propagates in free space, 
then strikes the plasma slab. The intensity, wavelength, number 
of cycles, the duration of pulse, the width $L_p$ of the plasma slab can 
be varied as desired. Accordingly the length of the computational domain,
and the number of computational cells $N_g$ are also adjusted.
We choose the laser wavelength $\lambda=800$ nm with $n_c = 4$-cycles, 
and the total pulse duration $\approx 30$ fs. 
The size of a computational cell is chosen as $\Delta = 200$ a.u. which
yields the PIC time step $\dtpic = 0.729248$ a.u. 
Length of the plasma is chosen as $L_p \approx 1.32\lambda$ 
with a plasma density $\rho/\rho_c \approx 0.136$. 
Temperature of ions are kept fixed in all simulations 
at $\Ti = 5$~eV while temperature of electrons are kept fixed 
at different values, e.g., $\Te = 5, 10, ..., 50, 100$~eV for a 
given laser intensity. 
The chosen value of $\Ti$, however, is found to have negligible 
effect on the overall results of collisional absorption. Above parameters 
are kept fixed during a simulation run unless mentioned explicitly. 
%
%
%
To simulate inverse bremsstrahlung absorption in 
presence of a laser field the Coulomb logarithm should not be same 
as Eq.\reff{Clg} for ordinary collisions, since it does not include the 
laser field parameters. Because there is no unique model 
of $\ln\Lambda$ in presence of laser field, we use a modified 
Coulomb logarithm \cite{kundu1,mulser1,pert95}
\begin{equation}
\label{Clg2}
\lam = 0.5 \ln \left[ 1+ \left(b_\mathrm{max}/b_{\perp}\right)^2 \right],
\end{equation}
where
$b_{max} = V_t/\max(\omega,\omega_p)$, $b_{\perp} = e^2/(4\pi\epsilon_0 m_e V_t^2)$, with $V_t^2 = v_{th,e}^2 + v_0^2$ as the total velocity. The effect of laser field is incorporated through the ponderomotive velocity 
$v_0 = e E_0/m_e\omega$. 
In the absence of laser field, $V_t = v_{th,e}$, $b_{max} = V_t/\omega_p = \lambda_D$ and $\lam$ in Eq.\reff{Clg2} becomes nearly equal to that
given in Eq.\reff{Clg}.

\begin{figure}[h]
\includegraphics[width=0.4\textwidth]{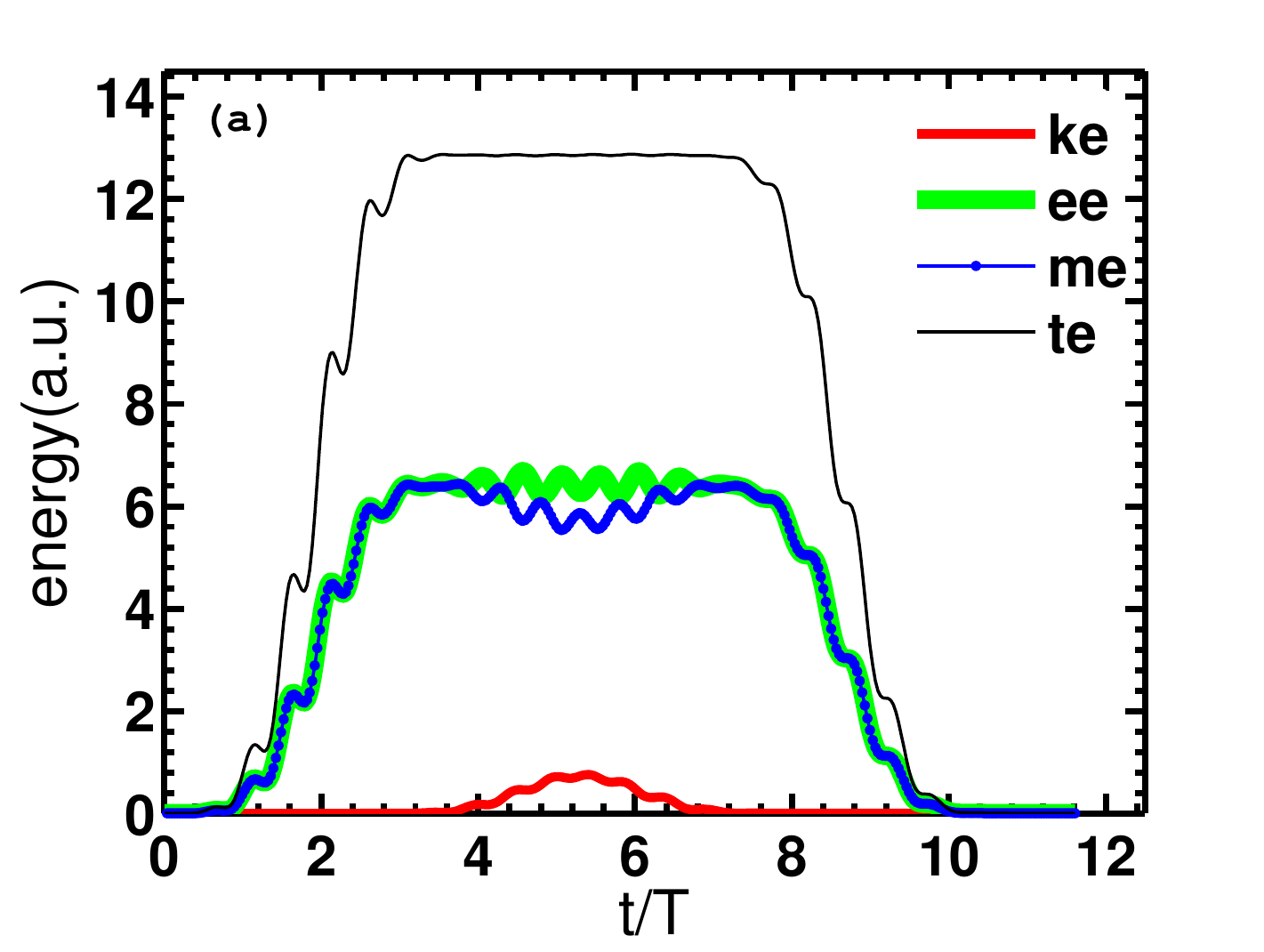}
\includegraphics[width=0.4\textwidth]{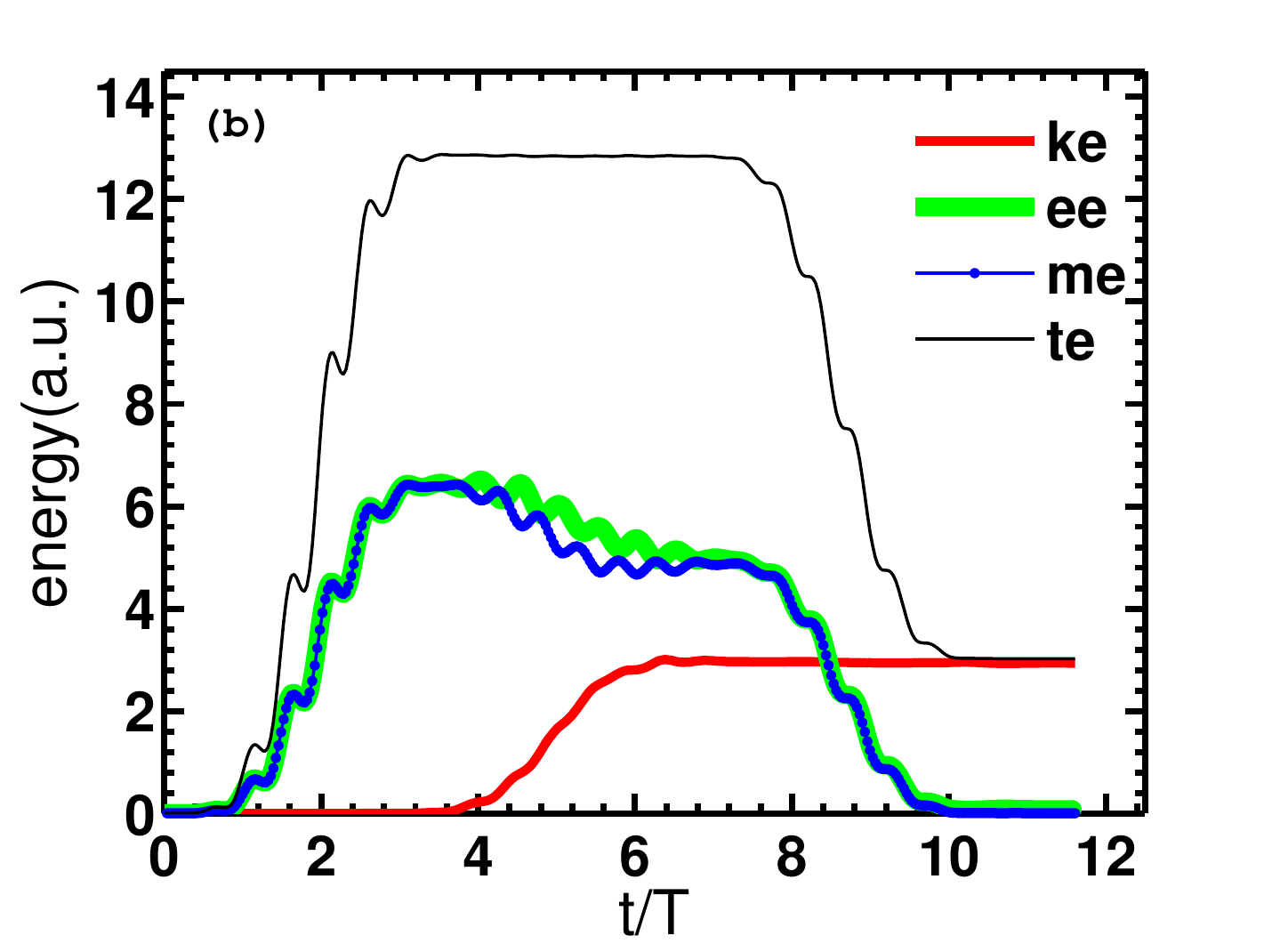}
\caption{Different energy profiles (kinetic energy $ke$, electric part $ee=\sum E_j^2\Delta/8\pi$, magnetic part $me=\sum B_j^2\Delta/8\pi$ of the electromagnetic energy, and the total energy $te=ke+ee+me$) 
vs normalized time $t/T$ for s-polarized light interacting with an
under-dense plasma. Without collision energy is not absorbed (in a) finally. 
With collision (in b) laser energy is absorbed, and transferred to 
the particle kinetic energies (see $ke\ne0$) in the end of the interaction.  
}
\label{tvsEng}
\end{figure}

From the simulation, we record total kinetic energy $ke$ gained by the particles, the electric part $ee=\sum_{1}^{N_g} E_j^2\Delta/8\pi$ and the magnetic part $me=\sum_{1}^{N_g} B_j^2\Delta/8\pi$ of the electromagnetic energy at every time step, giving the total energy $te=ke+ee+me$. 
Figure \ref{tvsEng} shows temporal variation of various energies 
at a given laser intensity $I_0 = 5\times 10^{14}\, \Wcmcm$ for the 
two cases: (a) without
collision, and (b) with collision between electrons and ions. 
In Figs.\ref{tvsEng} (a)-(b) for the initial time upto $t/T\approx 3$, all energies $ee, me$, and $te$ increase sharply since the laser pulse is entering
the simulation domain. For $t/T\approx3-4$ the value of $ee, me$ remains almost constant
and $te$ reaches a maximum because entire pulse has appeared in the 
simulation domain. The pulse strikes the plasma slab about $t/T\approx4$,
and only after this time, for $t/T = 4-7.5$, $ke$ first increases and then drops with 
the corresponding drop and increase in 
$ee$, and $me$ while $te$ remain conserved at the highest value. 
After $t/T\approx7.5$, values of 
$ee$, $me$ and $te$ sharply drop since the laser pulse is leaving the
finite simulation domain, and gets absorbed (artificially) in the 
right boundary.
The constant value of total energy, when the entire 
pulse is inside the computational box (for $t/T=3-7.5$), 
indicates conservation of energy in the simulation. 
In Fig.\ref{tvsEng}(a), without collision, $ke$ reaches a maximum value, 
and finally drops to zero before $t/T\approx7.5$. 
This is expected, because particles can not retain this energy, and 
finally give back to the electromagnetic fields (which is also evident 
from the corresponding drop and rise of $ee$ and $me$ between $t/T=4-7.5$), resulting no net absorption. 
However, when collision is taken into account [in Fig.\ref{tvsEng}(b)] 
$ke$ increases monotonically in time starting at $t/T \approx 4$ (with corresponding drop 
in $ee$ and $me$, meaning absorption of the pulse), and reaches a non-zero 
saturation value around $t/T = 6.8$ much before the pulse has left the simulation box. 
$ke$ does not drop to zero even after the pulse is over which
clearly shows that s-polarized light can be absorbed due to collisions and the 
laser energy can be transferred to the charge particles. 

\begin{figure}
\includegraphics[width=0.4\textwidth]{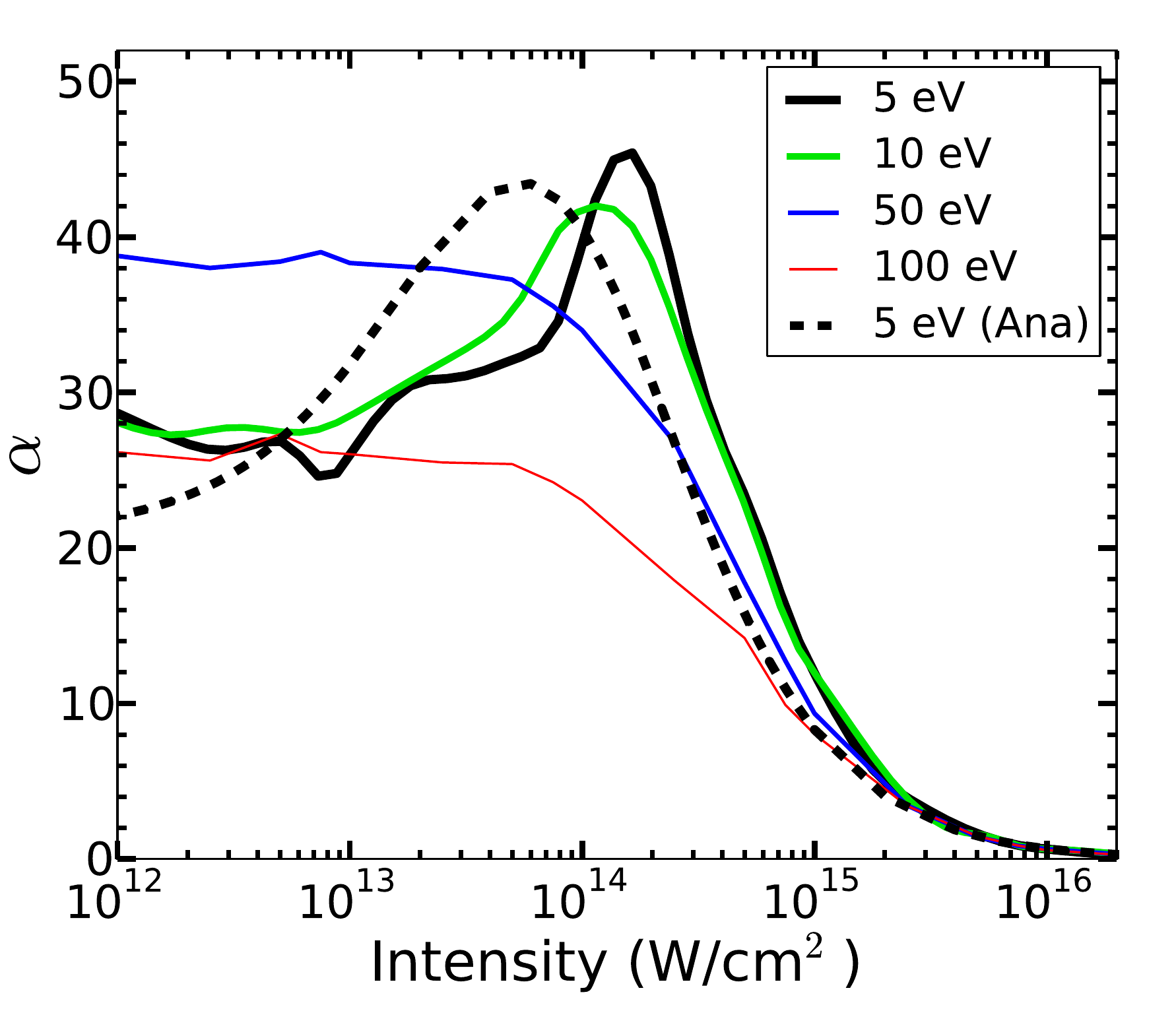}
\caption{Fractional absorption vs peak laser intensity, for 
s-polarized light interacting with an the under-dense plasma as 
in Fig.\ref{tvsEng} at $\Te = 5, 10, 50, 100$~eV (numerical, solid line) 
and at $\Te=5$~eV (analytical, dashed line, using Eq.\reff{alphaAbs}).}
\label{IvsAbs}
\end{figure}

We now find nature of collisional absorption by varying the 
intensity of the laser pulse for a given initial temperature $\Te$.
In reality, $\Te$ should also vary during the interaction. But 
our pulse being very short we assume it to be unchanged. The other parameters, such as plasma thickness $L_p$, plasma 
density $\rho$, ion temperature $\Ti$ are kept constant as above.

Figure~\ref{IvsAbs} shows fractional absorption $\alpha$, defined as the
ratio of the final kinetic energy retained in the particles to the 
maximum of $te$ (which is actually the total energy in the laser pulse), 
versus the peak intensity for $\Te = 5, 10, 50, 100$~eV (solid lines). 
It is seen that,
for higher temperatures $\Te>20$~eV,  $\alpha$ initially remains almost 
constant (or vary slowly) upto a certain value $I_c\approx 6\times 10^{13}\, \Wcmcm$ of the peak intensity,
and then decreases gradually for intensities $I_0>I_c$. This is the 
conventional result of collisional absorption reported in 
earlier works \cite{bornath,bor01,decker} with a $\ln\Lambda$ independent 
of the ponderomotive velocity $\v0=E_0/\omega$. However, at a lower 
$\Te<20$~eV, it is found that $\alpha$ initially increases with
the intensity, reaches a maximum value about an intensity $I_c$, then 
drops similar to the high temperature case. Such an anomalous behavior 
(initial increase followed by a drop) of fractional absorption versus the
laser intensity was reported experimentally with normally incident 
$s$-polarized light (of wavelengths 800~nm \cite{shalomBook} 
and 268~nm \cite{riley93} on an under-dense plasma with the peak 
absorption more than 30\%. Incorporating a total velocity dependent 
$\ln\Lambda$, in our EMPIC1D code assisted by Monte Carlo
binary collision we reproduce similar anomalous nature of collisional 
absorption in the low temperature regime. Our results indicate that 
fractional absorption due to collisional processes can be as high as 40\%
or even more for different plasma and laser parameters.  
%
%
For the shake of completeness, numerical results are compared 
with analytical estimates (dashed line in Fig.\ref{IvsAbs})
using a modified $\ln\Lambda$ as in Eq.\reff{Clg2} 
in the ballistic model \cite{mulser1,kundu1} of time-dependent 
electron-ion collision frequency  
\begin{equation}
\label{freqMulser2}
\nuei(t) = \frac{\omega_p^2 \lam}{\vos^3(t)} \left[\mathrm{erf} (u(t)) - \frac{2}{\sqrt{\pi}} u(t) e^{-u(t)^2}\right].
\end{equation}
Where $u(t) = \vos(t)/\sqrt{2} \vth$, and $\vos(t)$ is oscillation 
velocity of the electron in the laser field.
%
Averaging $\nuei(t)$ over a laser period leads to average 
$\nueibar$ and fractional absorption 
\begin{equation}
\label{alphaAbs}
\alpha = 1 - \exp({-2 \kappa_i L_p})
\end{equation}
of a continuous light of frequency $\omega$ 
in an under-dense plasma slab \cite{kruerBook,shalomBook} 
at normal incidence. 
Here $\kappa_i = (\rho/\rhoc) {\nueibar}/{\vg}$, and 
$\vg = c \sqrt{1 - \rho/\rhoc}$ is the group velocity of light. 
Analytical result using  
Eq.\reff{alphaAbs} (dashed line) at a lower temperature $\Te=5$~eV 
shows qualitative agreement with the EMPIC1D result, and confirms 
the anomalous nature of collisional absorption which was also reported 
by quantum and classical kinetic models \cite{kundu1,kunduPRE}. 
However, there are discrepancies between 
numerical and analytical results at higher temperatures,
which may be due to (i) time varying field experienced by particles, 
(ii) movement of ion back-ground to conserve momenta and 
energy during binary collisions in the numerical simulations as 
opposed to the analytical model where all particles experience same 
peak laser field $E_0$, and ions are considered stationary.

\section{Summary}\label{sec5}
Collisional absorption of \textit{s}-polarized laser light in a 
homogeneous, under-dense plasma is studied by a new 
particle-in-cell (PIC) simulation code considering one-dimensional 
slab-plasma geometry. To account for Coulomb collisions between charge 
particles a Monte Carlo (MC) binary collision scheme 
is used in the PIC code. 
For a given target thickness 
of a few times the wavelength of 800~nm laser 
fractional absorption of light due to Coulomb collisions 
is calculated at different electron temperature $\Te$ 
using a total velocity $v = \sqrt{\vth^2 + \v0^2}$ 
dependent Coulomb logarithm $\ln\Lambda(v)$. 
In the low temperature regime ($\Te\lesssim15$~eV) it is found that 
fractional absorption ($\alpha$) of light anomalously increases 
initially with increasing intensity $I_0$ up to a maximum value 
corresponding to an intensity $I_c$, and then it drops 
approximately obeying the conventional scenario, i.e., 
$\alpha \propto I_0^{-3/2}$ when $I_0>I_c$. 
Anomalous increase of $\alpha$ with $I_0$ was demonstrated in 
some earlier experiments \cite{shalomBook,riley93}, 
and recently explained by various models \cite{kundu1,kunduPRE} 
using total velocity dependent cut-offs. 
Here, we report anomalous nature of laser absorption by 
self-consistent PIC simulations assisted by Monte-Carlo collisions, 
thus bridging the gap between the models, simulations, and experiments.
%

%

%
\acknowledgments
The author would like to thank Anshuman Borthakur for the initial help in 
the Monte-Carlo simulations, 
the Plasma Science Society of India (PSSI) for providing 
partial financial support as a PSSI fellowship to carry out this work and
Sudip Sengupta for valuable suggestions.


\begin{thebibliography}{}
%
\bibitem{bach} D. R. Bach, D. E. Casperson, D. W. Forslund, S. J. Gitomer, P. D. Goldstone, A. Hauer, J. F. Kephart, J. M. Kindel, R. Kristal, G. A. Kyrala, K. B. Mitchell, D. B. van Hulsteyn, and A. H. Williams, \prl {\bf 50}, 2082 (1983).

\bibitem{teubner} U. Teubner, J. Bergmann, B. van Wonterghem, F. P. Sch\"afer, and R. Sauerbrey, \prl {\bf 70}, 794 (1993).

\bibitem{price} D. F. Price, R. M. More, R. S. Walling, G. Guethlein, R. L. Shepherd, R. E. Stewart, and W. E. White, 
\prl {\bf 75}, 252 (1995).

\bibitem{cerchez}
M. Cerchez, R. Jung, J. Osterholz, T. Toncian, O. Willi, P. Mulser, and H. Ruhl, \prl {\bf 100}, 245001 (2008).

\bibitem{rozmus} W. Rozmus and V. T. Tikhonchuk, 
\pra {\bf 46}, 7810 (1992).

\bibitem{dong} Q. L. Dong, J. Zhang, and H. Teng, \pre {\bf 64}, 026411 (2001).

\bibitem{kruerBook} William L. Kruer, \emph{The Physics of Laser Plasma Interactions, (Addison-Wesley, New York, 1988)}.
\bibitem{shalomBook} Shalom Eliezer, \emph{The Interaction of High-Power Lasers with Plasmas, (IOP Publishing, Bristol, 2002)}.
\bibitem{mulserBook} Peter Mulser and Dieter Bauer, \emph{High Power Laser-Matter Interaction, STMP 238 (Springer, Berlin, Heidelberg 2010)}.
%
\bibitem{gibbonBook} Paul Gibbon, \emph{Short Pulse Laser Interactions with Matter: An Introduction (Imperial College Press, 2005)}.


\bibitem{manes77} K. R. Manes, V. C. Rupert, J. M. Auerbach, P. Lee, and J. E.  Swain, Phys. Rev. Lett. {\bf 39}, 281 (1977).

\bibitem{mulserx0}
P. Mulser, D. Bauer, and H. Ruhl, Phys. Rev. Lett. {\bf 101}, 225002 (2008).
\bibitem{mulserx1} P. Mulser and M. Kanapathipillai, Phys. Rev. A {\bf 71}, 063201 (2005). 
\bibitem{mulserx2} P. Mulser, M. Kanapathipillai, and D. H. H. Hoffman, Phys. Rev. Lett. {\bf 95}, 103401 (2005).
\bibitem{kundu2} M. Kundu and D. Bauer, Phys. Rev. Lett. {\bf 96}, 123401 (2006).
\bibitem{kundu3} M. Kundu and D. Bauer, Phys. Rev. A {\bf 74}, 063202 (2006).
\bibitem{kundu4} M. Kundu, P. K. Kaw, and D. Bauer, Phys. Rev. A {\bf 85}, 23202 (2012).
\bibitem{kot03} I. Kostyukov and J. M. Rax, Phys. Rev. E {\bf 67}, 066405 (2003).

\bibitem{brunel} F. Brunel, Phys. Rev. Lett. {\bf 59}, 52 (1987).

\bibitem{mulserx3}
P. Mulser, S. M. Weng, and Tatyana Liseykinaa,
Phys. Plasmas {\bf 19}, 043301 (2012).

\bibitem{bauer07} 
%
D. Bauer and P. Mulser, Phys. Plasmas {\bf 14}, 023301 (2007).

\bibitem{hong06}
%
Hong-bo Cai, Wei Yu, Shao-ping Zhu, and Chun-yang Zheng,
Phys. Plasmas {\bf 13}, 113105 (2006).

\bibitem{bornath} Th. Bornath, D. Kremp, P. Hilse, and M. Schlanges, 
Journal of Physics: Conference series {\bf 11}, 180 (2005).

\bibitem{bor01} Th. Bornath, M. Schlanges, P. Hilse, and D. Kremp, \pre {\bf 64}, 26414 (2001).

\bibitem{riley93} D. Riley, L. A. Gizzi, A. J. Mackinnon, S. M. Viana, 
and O. Willi, \pre {\bf 48}, 4855 (1993).

\bibitem{sch97} L. Schlessinger and J. Wright, Phys. Rev. A {\bf 20}, 1934 (1979).
\bibitem{hil05} P. Hilse, M. Schlanges, Th. Bornath, and D. Kremp, Phys. Rev. E {\bf 71}, 056408 (2005). 
\bibitem{men13} 
J. T. Mendon\c{c}a, R. M. O. Galv\~{a}o, A. Serbeto, Shi-Jun Liang, and L. K. Ang, Phys. Rev. E {\bf 87}, 063112 (2013).
\bibitem{mol12} M. Moll, M. Schlanges, Th. Bornath, and V. P. Krainov, New Journal of Physics {\bf 14}, 065010 (2012).
\bibitem{pert72a} G. J. Pert, J. Phys. A {\bf 5}, 506 (1972).
\bibitem{pert75} G. J. Pert, J. Phys. B {\bf 8}, 3069 (1975). 
\bibitem{rand64} S. Rand, Phys. Rev. {\bf 136}, B231 (1964). 
\bibitem{weng95} Su-Ming Weng, Zheng-Ming Sheng, and Jie Zhang, \pre {\bf 80}, 56406 (2009).
\bibitem{silin} V. P. Silin, Sov. Phys. JETP {\bf 20}, 1510 (1965).
\bibitem{mulser1} P. Mulser and A. Saemann, Contrib. Plasma Phys. {\bf 37}, 211 (1997).
\bibitem{mulser2} P. Mulser and R. Schneider, J. Phys. A: Math. Theor. {\bf 42}, 214058 (2009).
\bibitem{mulser3} P. Mulser, F. Cornolti, E. B\'esuelle, and R. Schneider, \pre {\bf 63}, 16406 (2000).
\bibitem{kull01} H.-J. Kull and L. Plagne, \pop {\bf 8}, 5244 (2001). 
\bibitem{wesson} John Wesson, \emph{Tokamaks, (Oxford University Press, 2004)}. 
\bibitem{pert95} G. J. Pert, \pre {\bf 51}, 4778 (1995). 
 

\bibitem{rae92} S. C. Rae and K. Burnett, \pra {\bf 46}, 2077 (1992).
\bibitem{catto77} P. J. Catto and Th. Speziale, Phys. Fluids {\bf 20}, 167 (1977). 
\bibitem{kremp01} D. Kremp, Th. Bornath, P. Hilse, H. Haberland, M. Schlanges, and M. Bonitz, Contrib. Plasma Phys. {\bf 41}, 259 (2001).
\bibitem{brantov03} A. Brantov, W. Rozmus, R. Sydora, C. E. Capjack, V. Yu. Bychenkov et al., \pop {\bf 10}, 3385 (2003). 
\bibitem{skupsky} S. Skupsky, Physical review A {\bf 36}, 5701 (1987).

\bibitem{kundu1} M. Kundu, \pop {\bf 21}, 13302 (2014).
\bibitem{kunduPRE} M. Kundu, \pre {\bf 91}, 043102 (2015).
%


\bibitem{birdsall} C. K. Birdsall and A. B. Langdon, \emph{Plasma Physics via Computer Simulation, (McGraw Hill, New York, 1981)}.


\bibitem{hockney} R. W. Hockney and J. W. Eastwood, \emph{Computer Simulation using Particles, (IOP Publishing, Adam Hilger, New York, 1988)}.

\bibitem{sentoku} Y. Sentoku, K. Mima, Y. Kishimoto and M. Honda, Journal of Physical Society of Japan {\bf 67}, 4084 (1998);
Y. Sentoku, A.J. Kemp, Journal of Computational Physics {\bf 227}, 6846 (2008).


\bibitem{cadjan} M. G. Cadjan and M. F. Ivanov, 
J. plasma phys. {\bf 61}, 89 (1999).

\bibitem{takizuka} T. Takizuka and H. Abe, 
J. Comput. Phys. {\bf 25}, 205 (1977).

\bibitem{takizuka2}
T. Takizuka, Plasma Phys. Control. Fusion {\bf 59}, 034008 (2017); 
T. Takizuka , K. Shimizu , N. Hayashi , M. Hosokawa, and M. Yagi,
Nucl. Fusion {\bf 49}, 075038 (2009); 
%
T. Takizuka, Plasma Science and Technology {\bf 13}, 316 (2011).

\bibitem{sma} S. Ma, R. D. Sydora and J. M. Dawson,
Comput. Phys. Commun. {\bf 77}, 190 (1993).

\bibitem{manheimer} W. M. Manheimer, M. Lampe and G. Joyce,
J. Comput. Phys. {\bf 138}, 563 (1997).

\bibitem{oliphant} T. A. Oliphant and C. W. Nielson, 
Phys. Fluids {\bf 13}, 2103 (1970).

\bibitem{lichters} R. Lichters,  R. E. W. Pfund, and J. Meyer-ter-vehn, 
\emph{A parallel one dimensional relativistic electromagnetic particle-in-cell code for simulating laser plasma interaction, 
Max~Planck~Institute, Garching, Germany, http://www.lichters.net/work.html}.

\bibitem{harmut} H. Ruhl, \emph{Classical particle simulations, in: 
Introduction to Computational Methods in Many Body Physics, M. Bonitz, D. Semkat (Eds.), (Rinton Press, 2006)}; K. Germaschewski, W. Fox, N. Ahmadi, L. Wang, S. Abbott, H. Ruhl, A. Bhattacharjee,
\emph{The Plasma Simulation Code: A modern particle-in-cell code with load-balancing and GPU support, http://arxiv.org/abs/1310.7866v1}.

\bibitem{pukhov} A. Pukhov, J. Plasma Phys. {\bf 61}, 425-433 (1999). 


\bibitem{bowersVPIC} K. J. Bowers, B. J. Albright, L. Yin, B. Bergen, T. J. T. Kwan, 
Physics of Plasmas {\bf 15}, 055703 (2008). 

\bibitem{foncecaOSIRIS} R. A. Fonseca, L. O. Silva, F. Tsung, V. K. Decyk, W. Lu, C. Ren, W. B. Mori, S. Deng, S. Lee, T. Katsouleas, et al., \emph{OSIRIS: A three-dimensional,
fully relativistic particle in cell code for modeling plasma based accelerators, in: (Computational Science--ICCS 2002, Springer,
pp. 342 - 351)}.

\bibitem{nieterVORPAL} C. Nieter, J. R. Cary, 
Journal of Computational Physics {\bf 196}, 448 (2004). 

\bibitem{umeda} 
T. Umeda, Y. Omura, T. Tominaga, H. Matsumoto, 
Computer Physics Communications {\bf 156}, 73 (2003).
%

\bibitem{sullivan} D. M. Sullivan, 
\emph{Electromagnetic Simulation Using the FDTD Method, 2nd Edition, 
(Wiley-IEEE Press, 2013)}.

\bibitem{Ginzburg} 
V. L. Ginzburg, \emph{Propagation of electromagnetic waves in plasma,
(Pergamon Press, Oxford, 1970)}. 
%
\bibitem{truvnikov} 
B. A. Trubnikov, \emph{Particle Interactions in a Fully Ionized Plasma, in: Reviews of Plasma Physics, Vol. 1, M. A. Leontovich, Ed., (Consultants Bureau, New York, 1965)}. 
\bibitem{pert99} G.J. Pert, J Phys B {\bf 38}, 27 (1999). 
\bibitem{cohen2013} B. I. Cohen, A. M. Dimits, D. J. Strozzi, Journal of Computational Physics {\bf 234}, 33 (2013). 

\bibitem{spitzer} L.  Spitzer Jr., \emph{Physics of Fully Ionized Gases (Interscience Publishers, New York, 1956)}.

\bibitem{decker} C. D. Decker, W. B. Mori, J. M. Dawson, and T. Katsouleas, Phys. Plasmas {\bf 1}, 4043 (1994).











\end{thebibliography}
\end{document}